\documentclass{article}

    \PassOptionsToPackage{numbers, compress}{natbib}
 \usepackage[preprint]{neurips_2026}

\usepackage[utf8]{inputenc} 
\usepackage[T1]{fontenc}    
\usepackage{hyperref}       
\usepackage{url}            
\usepackage{booktabs}       
\usepackage{amsfonts}       
\usepackage{nicefrac}       
\usepackage{microtype}      
\usepackage{xcolor}         
\usepackage{amsmath,amssymb}
\usepackage{multirow}
\usepackage{colortbl}
\usepackage{array}
\usepackage{makecell}
\usepackage{algorithm}
\usepackage{algorithmic}
\usepackage{tabularx}
\usepackage{caption}
\usepackage{wrapfig}        
\DeclareMathOperator{\diag}{diag} 
\usepackage{tikz}
\usetikzlibrary{shapes, arrows.meta, positioning}

\usepackage{pifont} 

\newcommand{\yes}{\ding{51}}   
\newcommand{\no}{\ding{55}}    

\title{A 3GPP-Compliant Benchmark Dataset for RIS-Aided Beyond 5G Networks}

\author{%
 Pujitha Mamillapalli
    \\
  Department of Artificial Intelligence\\
  Indian Institute of Technology\\
 Hyderabad, India \\
  \And
  Pankaj Singh Rathour \\
  Department of Electrical Engineering \\
  Indian Institute of Technology\\
 Hyderabad, India \\
  \AND
  Abhinav Kumar \\
  Department of Electrical Engineering \\
  Indian Institute of Technology\\
 Hyderabad, India \\
}

\begin{document}

\maketitle

\begin{abstract} Reconfigurable Intelligent Surfaces (RIS) are emerging as a key technology for programmable wireless environments in the beyond the fifth generation (B5G) networks. However, data-driven RIS research remains bottleneck by the lack of standardized, high-fidelity and open-source datasets. In this paper, we introduce a large-scale 3GPP TR 38.901-compliant dataset for RIS-aided millimeter wave (mmWave) networks, that considers severe path loss, blockage sensitivity, and spatial channel sparsity make the RIS assistance more impactful. The dataset spans various canonical 3GPP  deployment scenarios across 20 controlled variants, capturing diverse user densities, fading conditions, and blockage regimes. Uniquely, every sample includes oracle RIS phase configurations obtained via a globally optimal brute-force codebook search, providing gold-standard supervision labels that are absent from any existing public dataset. Rich multi-task annotations comprising full channel state information (CSI), per-link channel decomposition, optimal phase matrices, and channel quality index (CQI) labels support a broad range of machine learning paradigms and downstream tasks, including phase optimization, channel estimation, and interference management. As the primary benchmark task, we introduce a novel CSI-to-CQI mapping that frames RIS-aided link-quality prediction as a scalable scalar classification problem, thereby avoiding the exponential output complexity of the direct phase vector prediction. We have evaluated this mapping against state-of-the-art architectures under in-distribution, out-of-distribution, and real-world hardware measurement conditions. Our dataset provides a reproducible, extensible, and community-ready foundation to accelerate data-driven research in RIS-aided B5G networks. \end{abstract}

\section{Introduction}

The convergence of massive device proliferation, immersive applications, and stringent latency requirements is pushing wireless infrastructure past its conventional limits in beyond the fifth generation (B5G) networks~\cite{itu2023imt2030}. Reconfigurable intelligent 
surfaces (RIS) — planar arrays of independently tunable, passive or 
semi-passive electromagnetic elements — have emerged as a key technology. Unlike traditional wireless equipment, it can program how signals travel in real time — without needing power amplifiers or other complex radio hardware.~\cite{basar2019wireless}. By programming the signals direction, a properly set up RIS can reach areas with poor coverage, reduce signal interference between users and improve overall signal strength. Thus the RIS can transforms an otherwise uncontrollable stochastic channel into a deterministic and programmable resource at a fraction of the energy cost of 
active alternatives~\cite{renzo2020smart, huang2019reconfigurable}. However, optimizing RIS configurations remains computationally challenging, particularly in large-scale deployment scenarios, where the high-dimensional, non-convex nature of the problem significantly increases complexity~\cite{wu2019beamforming}. Classical model-based iterative solvers — alternating optimization~\cite{guo2020weighted}, successive convex 
approximation~\cite{huang2019reconfigurable}, and manifold 
optimization~\cite{yu2019maris} — offer convergence guarantees but scale poorly with high computation cost rendering them impractical under blockage. These limitations have catalyzed a growing research on machine learning (ML)-driven approaches, spanning deep reinforcement learning~\cite{feng2020deep, taha2021deep}, supervised codebook prediction~\cite{yang2020deep, hu2021iterative}, and unsupervised learning~\cite{CNN}, all promising real-time inference and robustness to model mismatch~\cite{CNN}. However, the progress is 
severely hampered by a fundamental bottleneck: the absence of realistic, standardized and large-scale datasets~\cite{alkhateeb2019deepmimo}. Existing works predominantly rely on privately generated, scenario-specific datasets built on simplified channel models that neither conform to 3GPP propagation standards~\cite{3gpp38901} nor provide sufficient diversity to support generalization. Furthermore, the absence of quality  labels foreclose rigorous evaluation of supervised and imitation-learning approaches, collectively imposing a substantial hindrance on the field.

To address these limitations, we introduce a large-scale 3GPP TR~38.901-compliant dataset for RIS-aided mmWave networks as a community benchmark. Our dataset spans three canonical deployment scenarios: Urban Macro (UMa), Urban Micro (UMi), and Indoor Hotspot (InH), across 20 controlled variants capturing diverse user densities, fading levels, and blockage probabilities per 3GPP TR~38.901~\cite{3gpp38901}. Each data sample carries channel state information (CSI), per-link channel decomposition and most distinctively, RIS phase labels obtained via a globally optimal brute-force codebook search~\cite{wu2019beamforming, bjornson2020reconfigurable}. These labels provide standard supervision targets which are  not available in many existing dataset. 
We benchmark our dataset on a novel CSI-to-channel quality indicator (CQI) mapping task using a suite of state-of-the-art architectures under in-distribution, out-of-distribution, and 
real-world measurement conditions~\cite{mamillapalli2026, brisc2025}, 
establishing reference performance numbers for the community.

The contributions of this work are as follows:
\begin{itemize}
    \item \textbf{Large-scale 3GPP-compliant dataset.} A dataset generated 
    using TR~38.901~\cite{3gpp38901} that spans various scenarios across 20 controlled variants with diverse fading models, user 
    densities, and blockage regimes.

    \item \textbf{Richly annotated dataset.} Optimal RIS phase configurations obtained via exhaustive codebook search provide gold-standard supervision, complemented by comprehensive multi-task annotations—including full CSI with per-link decomposition, beamforming vectors, optimal phase matrices, and CQI labels—supporting a wide range of supervised, unsupervised, and reinforcement learning tasks.

   \item \textbf{Novel CSI-to-CQI mapping framework.} We introduce a novel framework that models RIS phase optimization as a direct mapping from CSI to a scalar CQI, avoiding the exponential output complexity of phase vector prediction while yielding an operationally meaningful link-adaptation target.

    \item \textbf{Benchmarking and generalization analysis. }Comprehensive evaluation of diverse learning paradigms—including deep neural networks, sequence modeling approaches, and classical machine learning methods—with a focused analysis of model generalization under in-distribution, out-of-distribution, and real-world hardware measurement conditions.

\end{itemize}
\begin{table*}[t]
\centering
\caption{Comparison of related works on RIS datasets and simulation frameworks.
GBSCM\,=\,geometry-based stochastic channel model, RT\,=\,Ray Tracing,
CM = Channel Modeling, RW Meas. = Real World Measurements and  Avail. = Availability.}
\label{tab:related_work}
\renewcommand{\arraystretch}{1.2}
\resizebox{\textwidth}{!}{%
\begin{tabular}{%
  l
  l
  c
  c
  c
  c
  c
  c
  c
}
\toprule
\textbf{Work} &
\textbf{Channel Model} &
\textbf{Frequency} &
\makecell{\textbf{Multi-} \\ \textbf{scenario}} &
\makecell{\textbf{3GPP} \\ \textbf{TR 38.901}} &
\makecell{\textbf{RIS } \\ \textbf{CM}} &
\makecell{\textbf{Oracle} \\ \textbf{Labels}} &
\makecell{\textbf{RW} \\ \textbf{Meas.}} &
\makecell{\textbf{Dataset } \\ \textbf{Avail.}} \\

\midrule
\multicolumn{9}{l}{\textit{Stochastic channel models}} \\[2pt]

QuaDRiGa \cite{jaeckel2014quadriga}            & GBSCM            & Sub-6/mmWave  & \yes & \yes & \no & \no & \no & \no \\
COST~2100 \cite{liu2012cost}                   & GBSCM            & Sub-6\,GHz    & \yes & \no & \no & \no & \no & \no \\

\midrule
\multicolumn{9}{l}{\textit{Ray tracing frameworks}} \\[2pt]

DeepMIMO \cite{alkhateeb2019deepmimo}          & RT               & mmWave/sub-6  & \yes & \no & \no & \no & \no & \yes \\
Raymobtime \cite{klautau2018raymobtime}         & RT + traffic sim & 60\,GHz       & \no & \no & \no & \no & \no & \yes \\
Wireless InSite \cite{alkhateeb2015compressed} & RT (commercial)  & Configurable  & \yes & \no & \no & \no & \no & \no \\
Bayraktar et al.\ \cite{bayraktar2024}         & RT               & mmWave        & \no & \no & \yes & \no & \no & \yes \\
From Lab to DT \cite{lab2dt2025}               & RT + calibration & mmWave        & \no & \no & \yes & \no & \no & \no \\
SionnaRT \cite{sionna2022, sionnart2023}                   & RT + 3GPP        & Sub-6/mmWave  & \yes & \yes & \no & \no & \no & \yes \\



\midrule
\multicolumn{9}{l}{\textit{Real-world RIS measurements}} \\[2pt]

Tewes et al.\ \cite{tewes2023}                & Measured         & sub-6        & \no & \no & \yes & \no & \yes & \yes \\

Mamillapalli et al.\ \cite{mamillapalli2026}   & Measured         & sub-6/mmWave   & \no & \no & \yes & \no & \yes & \no \\

BRISC \cite{brisc2025}                         & Measured         & sub-6        & \no & \no & \yes & \no & \yes & \yes \\

\midrule
\textbf{Ours}                          & 3GPP TR~38.901   & mmWave       & \yes & \yes & \yes & \yes & \no & \yes \\

\bottomrule
\end{tabular}%
}
\end{table*}

\section{Related Work}
In this section, we survey the  prior literature on RIS-aided systems, including simulation datasets, real-world measurements, and learning-based phase optimization, and highlight limitations in existing datasets that motivate the proposed approach.

Machine learning (ML) methods are increasingly replacing classical iterative solvers for real-time RIS phase configuration. Prior work spans supervised codebook prediction~\cite{yang2020deep, hu2021iterative}, convolutional neural network (CNN)-based feature learning~\cite{elbir2020cnn}, deep reinforcement learning (DRL)~\cite{feng2020deep, taha2021deep}, and transformer architectures~\cite{vaswani2017attention, lin2023transformer}. Classical optimization methods are theoretically grounded but suffer from high computational cost and large inference latency due to iterative updates~\cite{wu2019beamforming, guo2020weighted}. This limits their use in large-scale real-time systems. Learning-based approaches enable fast inference but introduce new challenges. DRL methods are sample-inefficient and often unstable in high-dimensional action spaces~\cite{taha2021deep}. Unsupervised methods remove the need for labels but lack guarantees of convergence to globally optimal solutions in non-convex settings~\cite{sun2017learning, huang2020holographic}. Supervised methods are stable and efficient but depend on high-quality oracle or near-optimal labels. Despite this progress, most works rely on privately generated single-scenario datasets without standardized channel models, 3GPP compliance, or oracle labels. This limits fair comparison and weakens conclusions on generalization to realistic deployments.

\paragraph{Existing Datasets: }Datasets such as DeepMIMO~\cite{elbir2021deep}, Raymobtime~\cite{klautau2018raymobtime}, Wireless InSite~\cite{basar2019wireless}, and the RT dataset~\cite{bayraktar2024} are typically tied to specific scene geometries, often generating deterministic channel realizations for fixed environments rather than samples from calibrated stochastic models. Geometry-based stochastic channel models (GBSCMs), including QuaDRiGa~\cite{jaeckel2014quadriga} and COST 2100~\cite{liu2012cost}, provide greater statistical diversity and partial alignment with 3GPP models, but are primarily designed for conventional multiple-input multiple-output (MIMO) and focus on composite end-to-end channels, with limited support for explicit transmitter–RIS and RIS–receiver decomposition. Standards-aligned simulators such as Sionna~\cite{sionna2022} and SionnaRT~\cite{sionnart2023} implement TR 38.901 and move toward 3GPP-compliant data generation; however, as summarized in Table~\ref{tab:related_work}, publicly available datasets based on these tools remain limited in multi-scenario coverage, RIS specific channel decomposition, and oracle labels.

Real-world RIS measurement datasets remain limited in scope and scale. Prior works report indoor and outdoor measurements across sub-6,GHz and mmWave bands, demonstrating measurable SNR and coverage gains under practical hardware conditions~\cite{tewes2023, brisc2025, weinberger2024validating, trichopoulos2022design, santoro2024open, sang2023measurement, sang2023multi, pei2021ris, mamillapalli2026}. However, these datasets are typically small, limited to a few scenarios and frequencies, and show limited alignment with 3GPP standards. They also lack oracle phase labels, restricting their use for systematic machine learning benchmarking and generalization studies.

Table~\ref{tab:related_work} highlights three recurring limitations in existing datasets: \textit{(i) scenario diversity}—most datasets focus on a single environment, limiting systematic evaluation; \textit{(ii) 3GPP compliance}—few RIS-specific datasets are generated under TR 38.901-aligned propagation settings at 28,GHz; and \textit{(iii) RIS channel modeling}—limited availability of datasets with explicit per-link transmitter–RIS and RIS–receiver decomposition required for phase optimization. The proposed dataset is designed to address these aspects while additionally providing oracle phase labels and standardized open-source splits for reproducible benchmarking.
\section{Preliminaries}
\label{sec:Background}
\paragraph{System Model. }In this paper, we consider a downlink MIMO RIS-aided wireless system comprising a base station (BS) equipped with $N$ antennas, a passive RIS with $L^2$ reflecting elements, and $K$ single-antenna users. The RIS is modeled by a diagonal phase-shift matrix $\boldsymbol{\theta} = \left[e^{j\theta_1}, \dots, e^{j\theta_{L^2}}\right]$, where $\theta_l (\in [0, \pi]) \subseteq \mathcal{S}$. Here, $\mathcal{S}$ denotes a finite discrete phase codebook shared across all elements. Let $\mathbf{H}_{BR} \in \mathbb{C}^{L^2 \times N}$ and $\mathbf{g}_{k} \in \mathbb{C}^{L^2 \times 1}$ denote the BS-to-RIS and RIS-to-user channels, respectively, and $\mathbf{h}_{d,k} \in \mathbb{C}^{1 \times N}$ denote the direct BS-to-user link. The received signal at the $k^{th}$ users is given by
\begin{equation}
y_k = \left(\mathbf{g}_{k}^{\dagger} \diag(\boldsymbol{\theta}) \mathbf{H}_{BR} + \mathbf{h}_{d,k} \right)\mathbf{w}_ks_k + I_K + \mathbf{n}.
\end{equation}
where $\mathbf{w}_k \in \mathbb{C}^{N \times 1}$ is the $k^{th}$ user beamforming vector , $s_k$ is the transmitted symbol vector to the $k^{th}$ user, $I_k$ is the interference from other users, and $n_k \sim \mathbb{CN}(\mathbf{0}, \sigma_n^2)$ denotes additive white Gaussian noise (AWGN); where $\mathbb{CN}(\cdot,\cdot)$ represents the complex Gaussian distribution with mean $0$ and noise variance $\sigma_n^2$.

\begin{figure}[t]
    \centering
    \includegraphics[width=\linewidth]{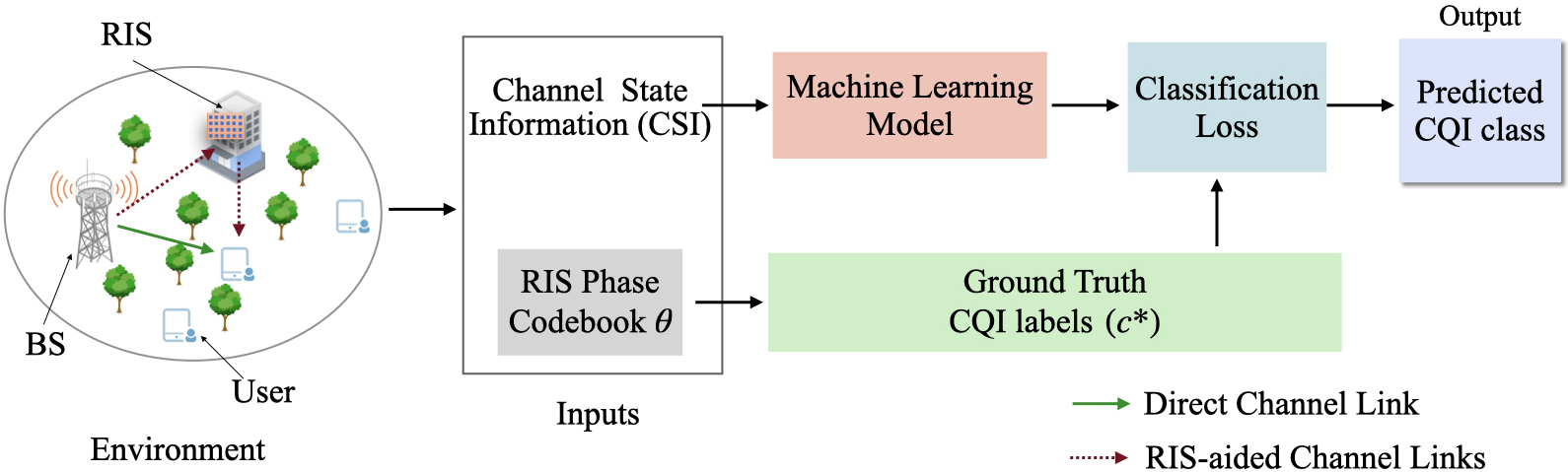}
    \caption{Overview of the proposed CSI-to-CQI mapping framework for RIS-aided mmWave networks.}
    \label{fig:task_framework}
\end{figure}

\paragraph{Performance Metrics: }In this paper, $\alpha$-mean throughput is considered as the performance metric, which is expressed as follows \cite{fairness}
\begin{align} \mathbb{T}_{\alpha}\!\left(k, \boldsymbol{\theta}, 
\boldsymbol{\mathcal{H}}\right)=\begin{cases} \displaystyle \left ({\frac {1}{|K|}\sum \limits _{k =1}^K (B * R_{k})^{1-\alpha } }\right)^{\frac {1}{1-\alpha }},&\text {$\alpha >0, \,\,\alpha \neq 1$,} \\ \displaystyle \left ({\prod \limits _{k =1}^K B * R_{k}}\right)^{\frac {1}{|K|}},&\text {$\alpha =1$,} \end{cases} 
\label{eq:throughput_fn}
\end{align}
where $\alpha$ is the fairness parameter, $B$ is the bandwidth, $K$ is the total number of UEs, and $R_k \triangleq f(\boldsymbol{\mathcal{H}}, \mathbf{\theta})$ is the achievable rate (Refer to supplementary material).

\paragraph{RIS Phase Optimization Task and Label Generation.}

Given the composite channel state information (CSI) 
$\boldsymbol{\mathcal{H}} = \{\mathbf{H}_{BR}, \mathbf{G}, \mathbf{H}_d\}$, 
comprising the BS-to-RIS, RIS-to-UE, and direct BS-to-UE links respectively, 
the optimal RIS phase configuration is obtained by solving:
\begin{equation}
\boldsymbol{\theta}^*(\boldsymbol{\mathcal{H}}) = 
\arg\max_{\boldsymbol{\theta} \in \mathcal{S}^{L^2}} 
\sum_{k=1}^{K} \mathbb{T}_{\alpha}\!\left(k, \boldsymbol{\theta}, 
\boldsymbol{\mathcal{H}}\right),
\label{eq:prob_formulation}
\end{equation}
where $\mathbb{T}_{\alpha}(k, \boldsymbol{\theta}, \boldsymbol{\mathcal{H}})$ 
denotes the $\alpha$-mean throughput of the $k$-th user (as in \eqref{eq:throughput_fn} under phase configuration 
$\boldsymbol{\theta}$~\cite{fairness}, $\mathcal{S}$ is the discrete phase 
codebook, and $L^2 = \varphi$ is the number of RIS elements. The exhaustive search 
over $\mathcal{S}^{\varphi}$ guarantees global optimality but incurs complexity 
$\mathcal{O}(|\mathcal{S}|^{\varphi})$, which grows exponentially with the number of 
elements.

Beyond computational intractability at inference time, directly predicting 
$\boldsymbol{\theta}^* \in \mathcal{S}^{\varphi}$ as a supervised learning target 
imposes an exponentially large structured output space, making the learning 
problem itself statistically and computationally prohibitive. To resolve both 
issues simultaneously, we propose mapping the oracle-optimal configuration to a 
scalar Channel Quality Indicator (CQI). Specifically, for each CSI realization 
$\boldsymbol{\mathcal{H}}$, the oracle first computes the globally optimal phase 
vector $\boldsymbol{\theta}^*(\boldsymbol{\mathcal{H}})$ via exhaustive search, 
evaluates the resulting sum throughput 
$\mathbb{T}_{\alpha}^*= \sum_{k}\mathbb{T}_{\alpha}(k, \boldsymbol{\theta}^*, 
\boldsymbol{\mathcal{H}})$, and assigns a discrete CQI label via the 
standardized quantization function~\cite{etsi138101CQI, 3gpp38901}:
\begin{equation}
c^*(\boldsymbol{\mathcal{H}}) = 
Q\!\left(\mathbb{T}^*(\boldsymbol{\mathcal{H}})\right), 
\qquad c^* \in \{1, \dots, \mathcal{C}\},
\label{eq:cqi_label}
\end{equation}
where $Q(\cdot): \mathbb{R}_{+} \rightarrow \{1,\dots,\mathcal{C}\}$ partitions 
the throughput axis into $\mathcal{C}$ discrete levels aligned with 3GPP CQI 
table definitions~\cite{3gpp38901, etsi138101CQI}, and $\mathcal{C}$ denotes the total number of CQI levels. This formulation collapses the output space from $|\mathcal{S}|^{L^2}$ to $\mathcal{C}$ classes, reducing exponential structured prediction to efficient scalar classification while producing a label that is directly actionable by the BS scheduler without post processing. The resulting 
supervised learning problem is defined over the dataset 
$\mathcal{D} = \{(\boldsymbol{\mathcal{H}}_i, c_i^*)\}_{i=1}^{m}$, where a 
model $f_{\boldsymbol{\Omega}}: \boldsymbol{\mathcal{H}} \mapsto 
\{1,\dots,\mathcal{C}\}$ with parameters $\boldsymbol{\Omega}$ learns the 
CSI-to-CQI mapping from oracle-generated labels. The overall pipeline is 
illustrated in Fig.~\ref{fig:task_framework}, where a ML model learns the CSI-to-CQI mapping from generated  oracle labels.
\paragraph{Learning Framework. } The CSI-to-CQI mapping is formulated as an empirical risk minimization (ERM) problem~\cite{goodfellow2016deep}. Given $m$ independently and identically distributed (i.i.d.)\ samples $\mathcal{D} = \{(\boldsymbol{\mathcal{H}}_i, c_i^*)\}_{i=1}^{m}$ drawn from an unknown channel distribution $\mathcal{P}_{\mathcal{H}}$, the predictor $f_{\boldsymbol{\Omega}}: \boldsymbol{\mathcal{H}} \rightarrow 
\{1, \dots, \mathcal{C}\}$ minimizing:
\begin{equation}
\hat{f} = \arg\min_{f \in \mathcal{F}}\; \frac{1}{m}\sum_{i=1}^{m} 
\ell\!\left(f(\boldsymbol{\mathcal{H}}_i), c_i^*\right),
\end{equation}
where $\ell(\cdot,\cdot)$ is the loss function~\cite{goodfellow2016deep, lecun2015deep}. Since oracle labels $c^*$ 
are globally optimal by construction, any residual risk 
$\mathcal{L}(\hat{f}) = \mathbb{E}[\ell(f(\boldsymbol{\mathcal{H}}), 
c^*(\boldsymbol{\mathcal{H}}))]$ reflects genuine model capacity or 
distributional mismatch rather than label suboptimality, providing a 
noise-free measure of the true performance ceiling.

\noindent
\begin{minipage}[t]{0.48\linewidth}
\section{Dataset Generation Pipeline}

In this section, we mention the detail pipeline for the dataset generation. The pipeline is identical across all 20 variants; only the channel configuration \texttt{cfg} varies between runs, ensuring that all inter-variant differences arise solely from controlled changes in propagation parameters. Algorithm~\ref{algo:pipeline} summarizes the end-to-end procedure; full implementation details, 3GPP TR~38.901 model parameters~\cite{3gpp38901}, are provided in supplementary material.

\textbf{Topology sampling: } For dataset generation, we assumed the system model as mentioned in Section \ref{sec:Background}. The BS and RIS are fixed at predetermined locations. UEs are distributed as a homogeneous Poisson point process (PPP) of density $\lambda_k$ (UEs/km$^2$) on a disc of radius $R$ centred at the BS, yielding $K \sim \mathrm{Poisson}(\lambda_k A)$ users within the user deployment area $A$. Inter-cell interference is modeled via a single-tier hexagonal grid with inter-site distance (ISD) per 3GPP TR~38.901~\cite{3gpp38901}; each 
interfering BS transmits at full power with path-loss computed under the 
same TR~38.901 model as the serving link. 


\textbf{Large-scale fading: }LOS/NLOS states are assigned via distance-dependent probability functions and path-loss with log-normal shadowing, and is computed as per 3GPP TR~38.901~\cite{3gpp38901}.
\end{minipage}%
\hfill
\begin{minipage}[t]{0.50\linewidth}
\begin{algorithm}[H]
\caption{Proposed Dataset Generation Pipeline}
\label{algo:pipeline}
\begin{algorithmic}[1]
\REQUIRE Configuration $\texttt{cfg}$ (fading, user density, topology, blockage, bandwidth, carrier frequency), codebook $\mathcal{S}$
\ENSURE Channels $\{\mathbf{H}_{BR}, \mathbf{G}, \mathbf{H}_D\}$, optimal phase $\boldsymbol{\theta}^*$, CQI label $c^*$
\STATE Sample UE positions and blockage states according to $\texttt{cfg}$
\STATE Compute large-scale fading (3GPP TR~38.901)
\STATE Generate small-scale channels $\mathbf{H}_{BR}, \mathbf{G}, \mathbf{H}_D$
\STATE Initialize $\mathbb{T}_{\alpha}^* \leftarrow -\infty$
\FOR{each $\boldsymbol{\theta} \in \mathcal{S}^N$}
    \STATE Compute effective channel:
    $\mathbf{H}_{\mathrm{eff}} = \mathbf{G}\,\mathrm{diag}(\boldsymbol{\theta})\,\mathbf{H}_{BR} + \mathbf{H}_D$
    \STATE Compute beamforming matrix $\mathbf{W}$ using MMSE-ZF beamforming
    \STATE Calculate the aggregate throughput of the system $\mathbb{T}_{\alpha}$ as in~\ref{eq:throughput_fn}
    \IF{$\mathbb{T}_{\alpha} > \mathbb{T}_{\alpha}^*$}
        \STATE $\mathbb{T}_{\alpha}^* \leftarrow \mathbb{T}_{\alpha}(\boldsymbol{\theta})$
        \STATE $\boldsymbol{\theta}^* \leftarrow \boldsymbol{\theta}$
    \ENDIF
\ENDFOR
\STATE Compute CQI label: $c^* \leftarrow Q(\mathbb{T}_{\alpha}^*)$
\STATE Store sample
\end{algorithmic}
\end{algorithm}
\end{minipage}


Channels $\mathbf{H}_{BR}$, $\mathbf{G}$, and $\mathbf{H}_D$ are generated 
per 3GPP TR~38.901~\cite{3gpp38901} under the variant-assigned fading model 
and stored in explicit per-link form. For each $\boldsymbol{\theta} \in 
\mathcal{S}^{L^2}$, the effective channel 
$\mathbf{H}_{\mathrm{eff}}(\boldsymbol{\theta}) = \mathbf{G}\,
\mathrm{diag}(\boldsymbol{\theta})\,\mathbf{H}_{BR} + \mathbf{H}_D$ is 
formed, a zero forcing (ZF) beamforming~\cite{alkhateeb2014channel} $\mathbf{W}(\boldsymbol{\theta})$ is calculated and 
$\mathbb{T}_\alpha(k,\boldsymbol{\theta})$ is evaluated~\cite{kebede2022precoding}. 
The optimal $\boldsymbol{\theta}^*$ is obtained via~\eqref{eq:prob_formulation}, and three oracle quantities are stored per 
sample: the optimal phase vector $\boldsymbol{\theta}^*$, the CQI label 
$c^* \in \{1,\dots,\mathcal{C}\}$, and the optimal aggregate $\alpha$-mean throughput $\mathbb{T}^*_\alpha$.

\textbf{Reproducibility.}  
All random number generators are explicitly seeded, and the pipeline executes on deterministic GPU kernels using 64-bit complex arithmetic throughout.

\paragraph{System Parameters.} The dataset targets mmWave RIS deployments at $28$,GHz over a $100 \times 100,\text{m}^2$ area, with a BS equipped with $N=36$ antennas, a $30 \times 30$ RIS, and $K=100$ users. UE locations are uniformly sampled within scenario-specific bounds to ensure spatial diversity. It comprises $200{,}000$ channel realizations across $20$ deployment variants ($10{,}000$ each), with each sample represented by a $250{,}000$-dimensional feature vector (real and imaginary parts separated). Detailed dataset visualizations and statistical summaries of the generated dataset are presented in the supplementary material.

\paragraph{Implementation. }All experiments were run on $2$ Nvidia-A$6000$ GPUs, each with $48~\text{GB}$ RAM. The dataset size is around $\sim300$ GB. Dataset and code are available online\footnote{Dataset is available at \url{https://www.kaggle.com/datasets/pujitham/3gpp-ri-dataset}}\footnote{Codes are available on \url{https://github.com/wirelessai941-web/Benchmarking_3GPP_RIS_mmWave_dataset}}.

\paragraph{Limitations. }Our dataset is simulation-based and omits hardware impairments such as phase noise and quantization errors, which may impact real-world performance. It further assumes static user locations and fixed blockage, excluding mobility and time-varying propagation dynamics. Future extensions will incorporate hardware-in-the-loop measurements with real RIS prototypes, mobility-aware channel evolution via 3GPP TR~38.901, and dynamic blockage modeling under user mobility.

\begin{figure}
    \centering
    \includegraphics[width=0.92\linewidth]{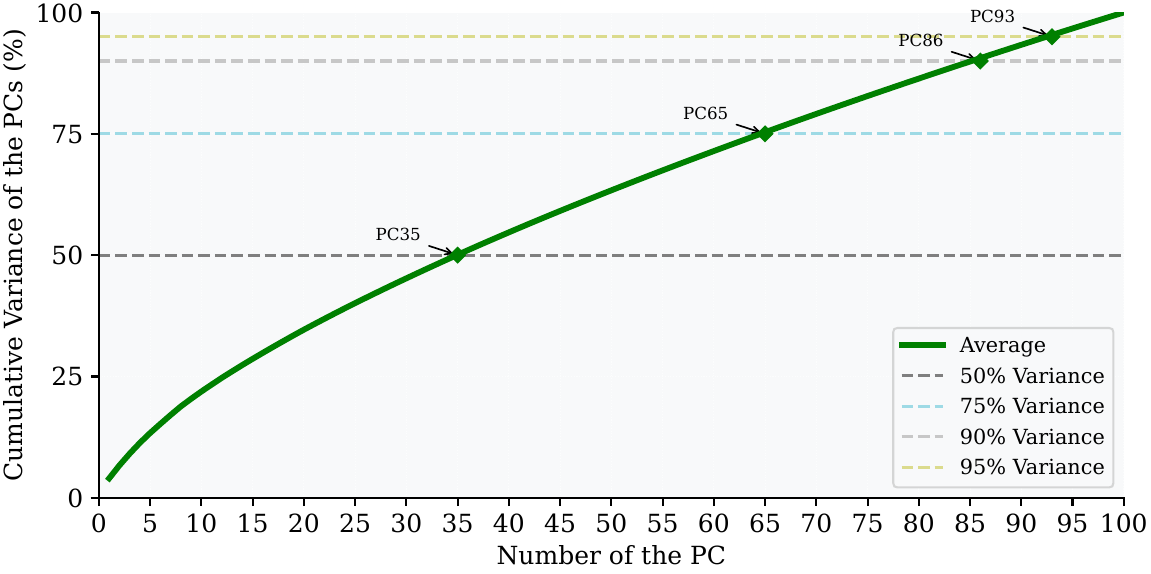}
    \caption{Cumulative variance explained as a function of the number of principal components (PCs). The solid curve denotes the average across all considered variants, while dashed horizontal lines indicate variance thresholds ($50\%, 75\%, 90\% ~\text{and}~ 95\%$). The intersection points highlight the minimum number of principle components (PC) required to achieve each threshold.}
    \label{fig:Cumm_EigenValues}
    \vspace{-0.25in}
\end{figure}
\section{Dataset Analysis}
This section provides a quantitative analysis of the dataset, focusing on its structure, variance distribution, and implications for learning. The dataset contains $200{,}000$ channel realizations across $20$ deployment variants, each represented by a $250{,}000$-dimensional feature vector. We analyze eigenvalue spectra to characterize variance concentration across principal components and its impact on learning and compression.

\subsection{Feature Variance Distribution via PCA}
To quantify intrinsic dimensionality, we perform the principal component analysis (PCA) on the sample covariance matrix and examine the cumulative explained variance ratio as $\mathcal{V}(r) = \frac{\sum_{i=1}^{r} \lambda_i}{\sum_{i=1}^{d} \lambda_i};$
where $\lambda_1 \geq \lambda_2 \geq \cdots \geq \lambda_d$ are the ordered eigenvalues. Fig.~\ref{fig:Cumm_EigenValues} shows the cumulative explained variance averaged across all variants. The curve increases smoothly without a clear elbow, indicating that the variance is not concentrated in a small subset of dominant components.

The first 35 components capture approximately 50\% of the total variance, representing the most significant large-scale channel variations. A substantial fraction of the variance is distributed across a wide range of components, with 65 and 86 components required to reach 75\% and 90\% variance, respectively. This near-linear accumulation reflects a broad spectrum of moderately informative directions, rather than a rapidly decaying eigenvalue profile. The remaining components contribute marginally, with a 95\% variance achieved at 93 components.

\paragraph{Implication—No Spectral Gap:} The gradual decay of the eigenvalues indicate the absence of a spectral gap, with variance distributed across many components rather than concentrated in a low-dimensional subspace. As a result, truncation-based dimensionality reduction is inherently lossy. This finding motivate models that preserve high-dimensional structure or learn task-aware compression. This behavior reflects realistic multi-cluster propagation and provides a challenging benchmark for evaluating robustness and generalization in data-driven RIS optimization.

\section{Evaluation Experiments}

In this section, we benchmark the dataset on the task defined in Section~\ref{sec:Background}: supervised CSI-to-CQI classification using oracle-optimal labels. The benchmark is designed to evaluate not only in-distribution generalization but also robustness to realistic distribution shifts — including noise model variation and real-world hardware-measured channels.

\paragraph{Benchmarking Models. }We evaluate a diverse set of models that span classical machine learning, deep learning, and sequence modeling paradigms, including feedforward neural networks (FNN) \cite{FNN}, CNN~\cite{li2021a}, eXtreme Gradient Boosting (XGBoost) \cite{chen2016xgboost}, logistic regression (LR)~\cite{Singh2024SVM}, support vector machines (SVM)~\cite{Singh2024SVM}, Transformer encoders \cite{vaswani2017attention}, and Mamba state space models \cite{gu2023mamba}. These models are selected for their complementary inductive biases: linear and tree-based methods provide strong baselines on compressed representations, FNN captures global feature interactions, CNN model local structure, and Transformer/Mamba architectures capture long-range dependencies—well aligned with our dataset distributed variance and absence of a spectral gap.

\begin{table*}[t]
\centering
\caption{Training and validation performance comparison. Best values per column are highlighted in \textbf{bold}. 
Acc\,=\,Accuracy, MAE\,=\,Mean Absolute Error, Params\,=\,Number of trainable parameters, GPU\,=\,GPU memory usage, Time/epoch\,=\,Training time per epoch.}
\label{tab:train_results}
\renewcommand{\arraystretch}{1.2}
\resizebox{\textwidth}{!}{%
\begin{tabular}{lcccccc}
\toprule
\textbf{Model} & \textbf{Train acc} $\uparrow$ & \textbf{Val acc} $\uparrow$ & \textbf{Params} $\downarrow$ & \textbf{MAE} $\downarrow$ & \textbf{GPU} $\downarrow$ & \textbf{Time/epoch} $\downarrow$ \\
\midrule
FNN~\cite{FNN}  & 85.6\%  & 85.4\%  & 0.45\,M  & 0.277  & 1.2~GB  & 0.5\,min \\
CNN~\cite{li2021a}  & 83.86\% & 82.54\% & 2.8\,M   & 0.295  & 1.8~GB  & ~\textbf{0.4\,min} \\
XGBoost~\cite{chen2016xgboost}  & \textbf{89.9\%} & \textbf{86.7\%}  & -- & \textbf{0.2456} & -- & 20\,min \\
LR~\cite{goodfellow2016deep}   & 68.43\% & 68.60\% & $\sim$5\,K & 0.412  & -- & 1\,min \\
SVM~\cite{Singh2024SVM}  & 81.92\% & 81.17\% & -- & 0.310  & -- & 25\,min \\
Transformer~\cite{lin2023transformer} & 84.8\% & 84.2\% & 33.3\,M & 0.309 & 3.9~GB & 1.2\,min \\
Mamba~\cite{gu2023mamba} & 85.5\% & 85.0\%  & 55.8\,M & 0.287 & 4.2\,GB & 1\,min \\
\bottomrule
\end{tabular}}
\vspace{2mm}

\raggedright

\end{table*}

\paragraph{Hyperparameters and Architecture.}
Raw CSI representations ($\boldsymbol{\mathcal{H}}$) are extremely high-dimensional ($\sim 2.5 \times 10^5$ features per sample), making direct end-to-end learning computationally expensive and prone to overfitting given $2 \times 10^5$ training samples~\cite{goodfellow2016deep}. To mitigate this, we extract 12 statistical descriptors—mean, standard deviation, maximum, minimum, median, energy, skewness, kurtosis, 25th/75th percentiles, range, and log-energy—from each of the three sub-channels, yielding a compact 36-dimensional feature vector. All features are standardized using training-set statistics. The FNN consists of four fully connected layers (512–512–256–128) with layer normalization and GELU activations, and a dropout of 0.2 is applied to the first three layers. A final linear layer maps to the output space. Training uses AdamW (learning rate $10^{-3}$, weight decay $10^{-4}$) with step decay (factor 0.7 every 10 epochs) for 100 epochs and batch size 64 under an 80:20 split. Kullback-Leibler (KL) divergence loss with batch-mean reduction is used for soft-label supervision, improving calibration over hard cross-entropy~\cite{NEURIPS2024}.  All models are trained under a standard empirical risk minimization (ERM) setting and are uniformly evaluated to establish consistent baseline performance.

Table~\ref{tab:train_results} highlights clear trade-offs between accuracy, training time, and model complexity. XGBoost achieves the best predictive performance, with the highest validation accuracy  and lowest mean absolute error (MAE)~\cite{elbir2020cnn}, but at the cost of significantly higher training time and no explicit control over model size. Among neural models, the FNN offers the most efficient balance, achieving competitive accuracy with low parameter count and fast training . The CNN trains fastest but delivers lower accuracy, indicating limited benefit from its convolutional structure. Transformer and Mamba models provide comparable accuracy  but incur substantially higher parameter counts and GPU memory usage, resulting in diminishing returns. Classical methods show mixed behavior: Logistic Regression is lightweight but under performs, while SVM improves accuracy at the expense of long training time. Overall, the results suggest that while XGBoost maximizes accuracy, compact neural models such as FNN offer a more favorable accuracy–efficiency trade-off for practical deployment.

\subsection{Test Datasets}

We evaluate generalization under four progressively challenging distribution 
shifts. \textbf{T1} is a held-out variant from the same generative process 
with a modified proportional fairness parameter $\alpha$ and increased noise 
variance, testing sensitivity to within-distribution parameter shifts. 
\textbf{T2} introduces impulsive bursty noise modeled as a Bernoulli-Gaussian 
mixture, evaluating robustness to heavy-tailed interference unseen during 
training~\cite{li2021a}. \textbf{T3} corrupts channels with temporally correlated auto regression AR(1) noise, 
introducing memory effects absent from the i.i.d.\ training distribution~\cite{li2021a}. 
\textbf{T4} uses hardware-measured RIS-aided channels from the BRISC 
dataset~\cite{brisc2025} at 5\,GHz, representing a joint shift in frequency, 
channel statistics, and hardware noise characteristics — the most severe 
sim-to-real evaluation. 

\paragraph{Train--Test Distribution Shift.}

\begin{wraptable}{r}{0.50\linewidth}
\vspace{-1em}
\centering
\caption{Train--test distribution shift via $D_{\mathrm{KL}}$, 
$\mathcal{W}_1$, and MMD. $\uparrow$ indicates larger shift from 
the training distribution, imp. = impairments.}
\label{tab:shift}
\renewcommand{\arraystretch}{1.2}
\resizebox{\linewidth}{!}{%
\begin{tabular}{llccc}
\toprule
\textbf{Set} & \textbf{Description} & 
$D_{\mathrm{KL}}$ $\uparrow$ & 
$\mathcal{W}_1$ $\uparrow$ & 
MMD $\uparrow$ \\
\midrule
T1 & AWGN        & 3.320          & 0.533          & 0.080 \\
T2 & Bursty         & 2.859          & 0.534          & 0.060 \\
T3 & Memory AR(1)        & 3.294          & 0.533          & 0.070 \\
T4~\cite{brisc2025} & Hardware imp.           & \textbf{8.741} & \textbf{1.872} & \textbf{0.284} \\
\bottomrule
\end{tabular}}
\vspace{-0.5em}
\end{wraptable}

We quantify the distributional distance between the training set and each 
test set using KL divergence $D_{\mathrm{KL}}$~\cite{NEURIPS2024}, 
Wasserstein-1 distance $\mathcal{W}_1$~\cite{NEURIPS2024}, and Maximum 
Mean Discrepancy (MMD) with an RBF kernel~\cite{li2021a}. As shown in Table~\ref{tab:shift}, T1--T3 exhibit comparable and moderate shift: T2 incurs slightly higher 
$D_{\mathrm{KL}}$ due to the heavy tail of the impulsive noise distribution, 
while T3 shows marginally lower MMD reflecting the smoother marginal 
perturbation of AR(1) correlated noise. T4 exhibits the largest shift across 
all three metrics by a substantial margin — more than $2.6\times$ that of any 
simulated test set — confirming that BRISC real-world 5\,GHz measurements 
occupy a fundamentally different region of feature space from the 
3GPP-simulated 28\,GHz training distribution. The ordering 
T1\,$\approx$\,T2\,$\approx$\,T3\,$\ll$\,T4 directly predicts the difficulty 
ranking observed in Table~\ref{tab:test_results}.

\begin{wraptable}{r}{0.5\linewidth}
\vspace{-10pt}
\centering
\caption{Test accuracy (\%) across T1--T4. Best in \textbf{bold}, FNN+FT: fine-tuning of the trained FNN model for 100 epochs.}
\label{tab:test_results}
\renewcommand{\arraystretch}{1.1}
\resizebox{\linewidth}{!}{%
\begin{tabular}{lcccc}
\toprule
\textbf{Model} & T1 & T2 & T3 & T4 \\
\midrule
LR~\cite{goodfellow2016deep}      & 25.2\% & 21.4\% & 23.1\% & 18.6\%\\
SVM~\cite{Singh2024SVM}     & 22.3\% & 28.7\% & 20.5\% & 12.4\% \\
XGB~\cite{chen2016xgboost}     & 39.5\% & 35.2\% & 37.8\% & 20.1\% \\
CNN~\cite{CNN}     & 52.3\% & 55.4\% & 58.9\% & 28.2\% \\
FNN~\cite{FNN}     & 61.4\% & 64.2\% & 66.8\% & 31.3\% \\
Transformer~\cite{lin2023transformer}  & 11.9\% & 10.0\% & 10.6\% & 9.8\% \\
Mamba~\cite{gu2023mamba}   & 25.6\% & 26.7\% & 25.1\% & 18.4\% \\
FNN+FT  & \textbf{90.8\%} & \textbf{91.3\%} & \textbf{87.8\%} & \textbf{78.4\%} \\
\bottomrule
\end{tabular}}
\vspace{-10pt}
\end{wraptable}

\paragraph{Results and Analysis.}
Table~\ref{tab:test_results} shows the test performance of the trained models across all conditions, along with an additional fine-tuning (FT) step applied to the FNN for 100 epochs. Classical models (LR, SVM, XGBoost) perform poorly, indicating limited capacity to capture the complex, high-dimensional structure of the RIS channels. Neural models significantly improve performance, with FNN achieving the strongest baseline results across simulated settings, suggesting that the extracted features are well-suited to dense representations. CNN shows moderate performance, while Transformer and Mamba under perform, indicating that higher model complexity does not guarantee better generalization without task-aligned inductive biases.
Under distribution shift, all models degrade, particularly in the sim-to-real setting (T4), confirming a substantial domain gap. Fine-tuning proves highly effective in mitigating this gap: FNN+FT achieves the best performance across all conditions, with significant gains on T4 using only 100 labeled samples. This improvement arises because the pretrained FNN captures general feature representations, which are efficiently adapted to target-domain statistics through fine-tuning.

\paragraph{Detailed Discussion: Generalization, Failure Modes, and Adaptation.}
Table~\ref{tab:test_results} reveals that model performance is limited not only under real-world shift (T4) but also across simulated conditions (T1–T3). Even without crossing the sim-to-real boundary, most models exhibit noticeable degradation across T1–T3, indicating that they fail to generalize across variations in noise statistics and channel dynamics within the same simulation domain. This suggests that the learned representations are highly sensitive to distributional changes and do not capture invariant features of RIS channels.

High-capacity architectures such as Transformer and Mamba perform particularly poorly, given the huge computation cost and lack of task-aligned inductive bias. Classical models (LR, SVM, XGBoost) also underperform due to limited representational capacity. While neural baselines such as FNN and CNN achieve relatively better results, their performance still drops consistently from T1 to T3 and further on T4, confirming that even these models struggle with cross-scenario generalization.

Fine-tuning significantly improves performance across all test sets. The strong results of FNN+FT indicate that the pretrained model captures useful but non-invariant features, which can be adapted to new distributions with minimal supervision. Notably, fine-tuning improves not only T4 (sim-to-real) but also T2 and T3, demonstrating that even simulated domain shifts require adaptation.

These findings are closely linked to the dataset design. Unlike prior datasets that focus on narrow, single-scenario settings, this dataset spans multiple deployment conditions with diverse channel characteristics. This diversity introduces substantial variability even within simulated data, making generalization inherently challenging. As a result, models trained on this dataset are forced to confront realistic distribution shifts rather than overfitting to a fixed environment.

Overall, the results show that state-of-the-art models for RIS applications do not generalize well across heterogeneous conditions without adaptation—even within simulation. This highlights a fundamental limitation of current learning approaches and underscores the need for domain generalization methods that can learn invariant representations across diverse scenarios. The proposed dataset provides a suitable and challenging benchmark for developing such approaches, enabling more robust and deployment-ready RIS optimization.

\section{Conclusion and Future Directions}

In this paper, we introduced a large-scale 3GPP TR 38.901-compliant dataset for RIS-aided mmWave networks that addresses key gaps in standardized channel modeling, per-link decomposition, oracle phase labels, and multi-scenario coverage with reproducible splits, along with a CSI-to-CQI mapping that reformulates RIS phase optimization as a scalable classification task. Extensive benchmarking shows that existing models struggle to generalize across both simulated and real-world distribution shifts, with performance degradation even within simulated settings, while fine-tuning provides only partial mitigation. Overall, our dataset offers a more diverse and realistic benchmark than prior single-scenario datasets, making it well-suited for evaluating robustness and domain generalization in RIS learning. However, the dataset remains simulation-based and does not capture hardware impairments or temporal mobility dynamics; future extensions will incorporate hardware-in-the-loop measurements and time-evolving channels to further improve realism and deployment relevance.

\paragraph{Boarder Impact.  } The proposed work can positively impact society by improving wireless coverage, energy efficiency, and reliability in future B5G networks, enabling applications such as smart cities, and autonomous systems. The dataset also supports reproducible research in RIS-aided learning systems. However, potential risks include misuse for enhanced surveillance or sensing, unequal access to advanced communication technologies.

{ \small
\bibliography{cite_NIPS26}
\bibliographystyle{ieeetr}
}
\newpage
\appendix

\newpage
\section*{NeurIPS Paper Checklist}

\begin{enumerate}

\item {\bf Claims}
    \item[] Question: Do the main claims made in the abstract and introduction accurately reflect the paper's contributions and scope?
    \item[] Answer: \answerYes{} 
    \item[] Justification: Our main contributions are presented in Section I, where we also provide the motivation for the dataset design and its key significance.

\item {\bf Limitations}
    \item[] Question: Does the paper discuss the limitations of the work performed by the authors?
    \item[] Answer: \answerYes{} 
    \item[] Justification: We explicitly discuss the limitations of our dataset in Section 4.

\item {\bf Theory assumptions and proofs}
    \item[] Question: For each theoretical result, does the paper provide the full set of assumptions and a complete (and correct) proof?
    \item[] Answer: \answerNA{}
    \item[] Justification: \justificationTODO{}

    \item {\bf Experimental result reproducibility}
    \item[] Question: Does the paper fully disclose all the information needed to reproduce the main experimental results of the paper to the extent that it affects the main claims and/or conclusions of the paper (regardless of whether the code and data are provided or not)?
    \item[] Answer: \answerYes{} 
    \item[] Justification: We have described the information needed to reproduce the main results in Section 4.

\item {\bf Open access to data and code}
    \item[] Question: Does the paper provide open access to the data and code, with sufficient instructions to faithfully reproduce the main experimental results, as described in supplemental material?
    \item[] Answer: \answerYes{}
    \item[] Justification: We provide open access to the complete dataset, accompanying metadata,
and all code necessary to reproduce the main experimental results. Detailed instructions and the access link are provided in Section 4. Dataset and code are available online\footnote{Dataset is available at \url{https://www.kaggle.com/datasets/pujitham/3gpp-ri-dataset}}\footnote{Codes are available on \url{https://github.com/wirelessai941-web/Benchmarking_3GPP_RIS_mmWave_dataset}}.


\item {\bf Experimental setting/details}
    \item[] Question: Does the paper specify all the training and test details (e.g., data splits, hyperparameters, how they were chosen, type of optimizer) necessary to understand the results?
    \item[] Answer: \answerYes{} 
    \item[] Justification: We explicitly discuss the the train and test details of our dataset in Section 6 and more detailed information is provided in the Supplementary material.

\item {\bf Experiment statistical significance}
    \item[] Question: Does the paper report error bars suitably and correctly defined or other appropriate information about the statistical significance of the experiments?
    \item[] Answer: \answerYes{} 
    \item[] Justification: We have reported the appropriate measures of statistical significance in the Section 6. 

\item {\bf Experiments compute resources}
    \item[] Question: For each experiment, does the paper provide sufficient information on the computer resources (type of compute workers, memory, time of execution) needed to reproduce the experiments?
    \item[] Answer: \answerYes{} 
    \item[] Justification: We have provided the details of computation resource in Section 4 implementation paragraph.
    
\item {\bf Code of ethics}
    \item[] Question: Does the research conducted in the paper conform, in every respect, with the NeurIPS Code of Ethics \url{https://neurips.cc/public/EthicsGuidelines}?
    \item[] Answer: \answerYes{} 
    \item[] Justification: Our research fully complies with the NeurIPS Code of Ethics. All dataset are generated via simulation setup as given in Section 4.

\item {\bf Broader impacts}
    \item[] Question: Does the paper discuss both potential positive societal impacts and negative societal impacts of the work performed?
    \item[] Answer: \answerYes{} 
    \item[] Justification: The boarder impact details are explicitly given in the conclusion section.
    
\item {\bf Safeguards}
    \item[] Question: Does the paper describe safeguards that have been put in place for responsible release of data or models that have a high risk for misuse (e.g., pre-trained language models, image generators, or scraped datasets)?
    \item[] Answer: \answerNA{} 
    \item[] Justification: \justificationTODO{}

\item {\bf Licenses for existing assets}
    \item[] Question: Are the creators or original owners of assets (e.g., code, data, models), used in the paper, properly credited and are the license and terms of use explicitly mentioned and properly respected?
    \item[] Answer: \answerYes{} 
    \item[] Justification: All external assets are properly acknowledged as detailed in the paper.

\item {\bf New assets}
    \item[] Question: Are new assets introduced in the paper well documented and is the documentation provided alongside the assets?
    \item[] Answer: \answerYes{} 
    \item[] Justification: We provide datasheets and detailed documentation for all newly introduced assets as part of the supplementary material.

\item {\bf Crowdsourcing and research with human subjects}
    \item[] Question: For crowdsourcing experiments and research with human subjects, does the paper include the full text of instructions given to participants and screenshots, if applicable, as well as details about compensation (if any)? 
    \item[] Answer: \answerNA{} 
    \item[] Justification: \justificationTODO{}

\item {\bf Institutional review board (IRB) approvals or equivalent for research with human subjects}
    \item[] Question: Does the paper describe potential risks incurred by study participants, whether such risks were disclosed to the subjects, and whether Institutional Review Board (IRB) approvals (or an equivalent approval/review based on the requirements of your country or institution) were obtained?
    \item[] Answer: \answerNA{} 
    \item[] Justification: \justificationTODO{}

\item {\bf Declaration of LLM usage}
    \item[] Question: Does the paper describe the usage of LLMs if it is an important, original, or non-standard component of the core methods in this research? Note that if the LLM is used only for writing, editing, or formatting purposes and does \emph{not} impact the core methodology, scientific rigor, or originality of the research, declaration is not required.
    \item[] Answer: \answerNA{} 
    \item[] Justification: \justificationTODO{}

\end{enumerate}

\end{document}